\documentclass{elsart}

\input epsf

\def\msun{M_{\odot}} 
\def\mpl{m_{{\rm Pl}}}
\def\delh{\delta_{{\rm H}}}

\begin{document}
\runauthor{A.~R.~Liddle}
\begin{frontmatter}

\title{Acceleration of the Universe} 
\author{Andrew R.~Liddle}
\address{Astronomy Centre, University of Sussex, Brighton BN1
9QJ, United Kingdom}

\begin{abstract}
The cosmological model best capable of fitting current observational
data features two separate epochs during which the Universe is
accelerating. During the earliest stages of the Universe, such
acceleration is known as cosmological inflation, believed to explain
the global properties of the Universe and the origin of
structure. Observations of the present state of the Universe strongly
suggest that its density is currently dominated by dark energy with
properties equivalent or similar to a cosmological constant. In these
lecture notes, I provide an introductory account of both topics,
including the possibility that the two epochs may share the same
physical description, and give an overview of the current status.
\end{abstract}

\end{frontmatter}

%%%%%%%%%%%%%%%%%%%%%%%%%%%%%%%%%%%%%%%%%%%%%%%%%%%%%%%%%%%%%%%%%%%%%%

\section{Overview} 

The original cosmological models assumed a Universe that underwent
deceleration during its entire evolution, corresponding to the slowing
of the expansion through gravitational deceleration. As cosmology has
developed, this assumption has been overturned at two separate
epochs. It is now widely believed that the Universe underwent a phase
of accelerated expansion, known as {\em cosmological inflation},
during some early stage of the Universe's expansion. This hypothesis
provides what is currently the best explanation for the observed
structures in the Universe, especially cosmic microwave background
anisotropies. In the present Universe evidence from many types of
observation, most prominently the apparent magnitude--redshift
relation of distant type Ia supernovae, suggests that the Universe is
presently accelerating, which can be explained by the existence of a
cosmological constant or an unusual form of matter sharing similar
properties, often referred to as {\em quintessence}.

The purpose of this article is to provide an introductory review of
these two topics, highlighting the ways in which the two phenomena may
be a consequence of the same underlying physical mechanism ---
domination of the Universe by the energy density of a scalar field. It
begins with an overview of inflation in the early Universe, with the
focus very much directed towards inflation as a theory of the origin
of structure. This is a particularly exciting time for proponents of
this mechanism, as microwave anisotropy measurements from the
BOOMERanG and MAXIMA experiments \cite{Boom,Max} have given strong
support to its basic predictions, opening the possibility of
high-precision testing in the future. Later in the article, the
current status of understanding of the present acceleration in the
context of scalar field domination is discussed, and challenges and
unsolved problems highlighted.

\section{A Hot Big Bang reminder}

\subsection{Overview}

To begin, I'll give a quick review of the big bang cosmology.  More
detailed accounts can be found in any cosmology textbook. The main
aim in this subsection is to set down the notation for the rest of the
article.

The standard hot big bang theory, by which I mean the description of
the Universe from a time of around one second onwards, is an extremely
successful one, passing some crucial observational tests of which I'd
highlight five.
\begin{itemize}
\item The expansion of the Universe.
\item The existence and spectrum of the cosmic microwave background
radiation.
\item The abundances of light elements in the Universe
(nucleosynthesis).
\item That the predicted age of the Universe is comparable to direct
age measurements of objects within the Universe.
\item That {\em given} the irregularities seen in the microwave
background by COBE, there exists a reasonable explanation for the
development of structure in the Universe, through gravitational
collapse.
\end{itemize}
In combination these are extremely compelling, and there is little
doubt that the physical framework for describing the Universe from one
second onwards is firmly in place, although the values of the various
parameters required to specify the model in detail remain uncertain.

However, there is a series of questions which the standard big bang
theory does not address. It does not predict the geometry of the
Universe, except insofar as to indicate that a spatially-flat
cosmology requires considerable fine-tuning of the initial
conditions. It does not predict the relative abundances of different
kinds of material --- baryons, radiation, dark matter, etc --- in the
present Universe; these are assumed. And, most importantly of all, it
does not offer an explanation for why the Universe was homogeneous to a
high degree of accuracy at early times, but with sufficient
irregularities within it to enable gravitational collapse to lead to
structures such as galaxies and galaxy clusters. In order to address
these questions, one must consider possible physical phenomena which
might have taken place in the very early Universe, at epochs so early
that the appropriate laws of physics are unknown. This raises the
exciting possibility that an accurate determination of the properties
of the present Universe might shed light on the physics of those early
stages.

\subsection{Equations of motion}

The hot big bang theory is based on the {\em cosmological principle},
which states that the Universe should look the same to all
observers. That tells us that the Universe must be homogeneous and
isotropic, which in turn tells us which metric must be used to
describe it. It is the Robertson--Walker metric
\begin{equation}
ds^2 = -dt^2 + a^2(t) \left[ \frac{dr^2}{1-kr^2} + r^2 \left(
	d\theta^2 + \sin^2 \theta \, d\phi^2 \right) \right] \,.
\end{equation}
Here $t$ is the time variable, and $r$--$\theta$--$\phi$ are (polar)
coordinates. The constant $k$ measures the spatial curvature, with $k$
negative, zero and positive corresponding to open, flat and closed
Universes respectively. If $k$ is zero or negative, then the range of
$r$ is from zero to infinity and the Universe is infinite, while if
$k$ is positive then $r$ goes from zero to $1/\sqrt{k}$. Many authors
rescale the coordinates to make $k$ equal to $-1$, $0$ or $+1$. The
quantity $a(t)$ is the scale-factor of the Universe, which measures
its physical size. The form of $a(t)$ depends on the properties of the
material within the Universe, as we'll see.

If no external forces are acting, then a particle at rest at a given
set of coordinates $(r,\theta,\phi)$ will remain there. Such
coordinates are said to be {\em comoving} with the expansion. One
swaps between physical (ie actual) and comoving distances via
\begin{equation}
\mbox{physical distance} = a(t) \times \mbox{comoving distance} \,.
\end{equation}

The expansion of the Universe is governed by the properties of
material within it. This can be specified\footnote{I follow standard
cosmological practice of setting the fundamental constants $c$ and
$\hbar$ equal to one.  This makes the energy density and mass density
interchangeable (since the former is $c^2$ times the latter). I shall
also normally use the Planck mass $\mpl$ rather than the gravitational
constant $G$; with the convention just mentioned they are related by
$G \equiv \mpl^{-2}$.} by the energy density $\rho(t)$ and the
pressure $p(t)$. These are often related by an equation of state,
which gives $p$ as a function of $\rho$; the classic examples are
\begin{equation}
\mbox{Radiation:}\; \; p = \frac{\rho}{3} \quad ; \quad
\mbox{Non-relativistic matter:} \; \;  p = 0 
\end{equation}
In general though there need not be a simple equation of state; for
example there may be more than one type of material, such as a
combination of radiation and non-relativistic matter, and certain
types of material, such as a scalar field, cannot be described by an
equation of state at all.

The crucial equations describing the expansion of the Universe are
\begin{eqnarray}
H^2 = \frac{8\pi}{3 \mpl^2} \, \rho - \frac{k}{a^2} + \frac{\Lambda}{3}
	\quad \quad & & \mbox{Friedmann equation} \\ 
\dot{\rho} + 3 H (\rho+p) = 0 \quad \quad \quad & & \mbox{Fluid equation}
\end{eqnarray}
where overdots are time derivatives and $H = \dot{a}/a$ is the Hubble
parameter. In this equation $\Lambda$ is the cosmological constant;
astronomers' convention is to write this as a separate term, though
physicists would typically be happier to consider it as part of the
density $\rho$.

These can also be combined to give
\begin{equation}
\frac{\ddot{a}}{a} = - \frac{4\pi}{3\mpl^2} \left( \rho + 3p \right) +
	\frac{\Lambda}{3} \quad
	\quad \mbox{Acceleration equation} 
\end{equation}
in which $k$ does not appear explicitly.

\subsection{Standard cosmological solutions}

When $k = \Lambda = 0$ the Friedmann and fluid equations can readily
be solved for the equations of state given earlier, leading to the
classic cosmological solutions
\begin{eqnarray}
&\mbox{Matter Domination~~~~~~} p = 0 : & \quad \rho \propto a^{-3} 
	\quad \quad a(t) \propto t^{2/3} \\
&\mbox{Radiation Domination~~~~~} p = \rho/3 : & \quad \rho \propto a^{-4} 
	\quad \quad a(t) \propto t^{1/2} 
\end{eqnarray}
In both cases the density falls as $t^{-2}$. When $k=0$ we have the
freedom to rescale $a$ and it is normally chosen to be unity at the
present, making physical and comoving scales coincide. The
proportionality constants are then fixed by setting the density to be
$\rho_0$ at time $t_0$, where here and throughout the subscript zero
indicates present value.

A more intriguing solution appears for domination by the cosmological
constant, namely
\begin{equation}
a(t) \propto \exp \left( Ht \right) \quad {\rm with} \quad H =
\sqrt\frac{\Lambda}{3} \,.
\end{equation}
This is equivalent to the solution for a fluid with equation of
state $p_\Lambda = -\rho_\Lambda$. The fluid equation then gives
$\dot{\rho}_\Lambda = 0$ and hence $\rho_\Lambda \equiv \Lambda/8\pi G$.

More complicated solutions can also be found for mixtures of
components.  For example, if there is both matter and radiation the
Friedmann equation can be solved using conformal time $\tau = \int
dt/a$, while if there is matter and a non-zero curvature term the
solution can be given either in parametric form using normal time $t$,
or in closed form with conformal time.

\subsection{Critical density and the density parameter}

The critical density $\rho_{{\rm c}}$ is defined as that giving a
spatially-flat geometry, $k = 0$, in the absence of a cosmological
constant. From the Friedmann equation, this implies
\begin{equation}
\rho_{{\rm c}}(t) = \frac{3 \mpl^2 H^2}{8\pi} \,.
\end{equation}
Densities are often measured as fractions of $\rho_{{\rm c}}$:
\begin{equation}
\Omega(t) \equiv \frac{\rho}{\rho_{{\rm c}}} \,.
\end{equation}
The quantity $\Omega$ is known as the density parameter, and can be
applied to individual types of material as well as the total density.

A similar definition can be employed for the cosmological constant,
giving
\begin{equation}
\Omega_\Lambda = \frac{\Lambda}{3H^2} \,,
\end{equation}
and when both density and cosmological constant are present the
condition for spatial flatness is $\Omega + \Omega_\Lambda = 1$.

The present value of the Hubble parameter is still not that well
known, and is parametrized as
\begin{equation}
H_0 = 100 h \; {\rm km \, s}^{-1} \, {\rm Mpc}^{-1} = \frac{h}{3000} 
	\, {\rm Mpc}^{-1} \,,
\end{equation}
where $h$ is normally assumed to lie in the range $0.5 \leq h \leq
0.8$. The present critical density is
\begin{equation}
\rho_{{\rm c}}(t_0) = 1.88 \, h^2 \times 10^{-29} \, {\rm g \, cm}^{-3} =
	2.77 \, h^{-1} \times 10^{11} \, \msun/(h^{-1} {\rm Mpc})^3 \,.
\end{equation}

\subsection{Characteristic scales and horizons}

The big bang Universe has two characteristic scales
\begin{itemize}
\item The Hubble time (or length) $H^{-1}$.
\item The curvature scale $a|k|^{-1/2}$.
\end{itemize}
The first of these gives the characteristic timescale of evolution of
$a(t)$, and the second gives the distance up to which space can be
taken as having a flat (Euclidean) geometry.  As written above they
are both physical scales; to obtain the corresponding comoving scale
one should divide by $a(t)$. The ratio of these scales gives a measure
of the total density; from the Friedmann equation we find
\begin{equation}
\sqrt{|\Omega + \Omega_\Lambda -1|} = \frac{H^{-1}}{a |k|^{-1/2}} \,.
\end{equation}

A crucial property of the big bang Universe is that it possesses {\em
horizons}; even light can only have travelled a finite distance since
the start of the Universe $t_*$, given by
\begin{equation}
d_{{\rm H}}(t) = a(t) \int_{t_*}^{t} \frac{dt}{a(t)} \,.
\end{equation}
For example, matter domination gives $d_{{\rm H}}(t) = 3t = 2H^{-1}$.
In a big bang Universe, $d_{{\rm H}}(t_0)$ is a good approximation to
the distance to the surface of last scattering (the origin of the
observed microwave background, at a time known as `decoupling'), since
$t_0 \gg t_{{\rm dec}}$.

\subsection{Redshift and temperature}

The redshift measures the expansion of the Universe via the stretching
of light
\begin{equation}
1+z = \frac{a(t_0)}{a(t_{{\rm emission}})} \,.
\end{equation}
Redshift can be used to describe both time and distance. As a time, it
simply refers to the time at which light would have to be emitted to
have a present redshift $z$. As a distance, it refers to the {\em
present} distance to an object from which light is received with a
redshift $z$.  Note that this distance is not necessarily the time
multiplied by the speed of light, since the Universe is expanding as
the light travels across it.

As the Universe expands, it cools according to the law
\begin{equation}
T \propto \frac{1}{a} \,.
\end{equation}
The expansion preserves the thermal form, in the absence of
interactions.  In its earliest stages the Universe may have been
arbitrarily hot and dense.

\subsection{The history of the Universe}

Presently the matter content of the Universe is dominated by
non-relativistic matter, but because radiation reduces more quickly
with the expansion, this implies that at earlier times the Universe
was radiation dominated. During the radiation era temperature and time
are related by
\begin{equation}
\frac{t}{1 \, {\rm sec}} \simeq 
    \left(\frac{10^{10} \, {\rm K}}{T} \right)^2 \,.
\end{equation}
The highest energies accessible to terrestrial experiment, generated
in particle accelerators, correspond to a temperature of about
$10^{15} \, {\rm K}$, which was attained when the Universe was about
$10^{-10} \, {\rm sec}$ old. Before that, we have no direct evidence
of the applicable physical laws and must use extrapolation based on
current particle physics model building. After that time there is a
fairly clear picture of how the Universe evolved to reach the present,
with the key events being as follows:
\begin{itemize}
\item $10^{-4}$ seconds: Quarks condense to form protons and neutrons.
\item 1 second: The Universe has cooled sufficiently that light nuclei
are able to form, via a process known as {\bf nucleosynthesis}.
\item $10^4$ years: The radiation density drops to the level of the
matter density, the epoch being known as {\bf matter--radiation
equality}. Subsequently the Universe is matter dominated.
\item $10^5$ years: Decoupling of radiation from matter leads to the
formation of the microwave background. This is more or less coincident
with recombination, when the up-to-now free electrons combine with the
nuclei to form atoms.
\item $10^{10}$ years: The present.
\end{itemize}
A more precise time-line is given in many textbooks,
e.g.~Ref.~\cite{LLbook}.

\section{The inflationary cosmology}

That inflation can resolve the classic initial conditions of the Hot
Big Bang model --- the horizon, flatness and monopole problems --- is
well documented (see e.g.~Ref.~\cite{Guth,KT}) and I will not repeat
the discussion here. Instead I will proceed directly to the definition
of inflation, and an explanation of the inflationary mechanism for the
origin of structure.

\subsection{The definition of inflation}

Inflation is defined to be any epoch during which the Universe is
accelerating, $\ddot{a} > 0$, with respect to cosmic time. We can
rewrite this in several different ways
\begin{equation}
\ddot{a} > 0 \quad \Longleftrightarrow \quad  \frac{d(H^{-1}/a)}{dt} < 0 
  \quad \Longleftrightarrow \quad  p < - \frac{\rho}{3} \,.
\end{equation}
The second of these is the most useful, because it has the most direct
geometrical interpretation. It says that the Hubble length, as
measured in comoving coordinates, {\em decreases} during inflation. At
any other time, the comoving Hubble length increases. This is the key
property of inflation; although typically the expansion of the
Universe is very rapid, the crucial characteristic scale of the
Universe is actually becoming smaller, when measured relative to that
expansion.

Quite a wide range of behaviours satisfy the inflationary
condition. The classic one is de Sitter expansion, which arises when
the Universe is dominated by a cosmological constant, which we saw
earlier gives $a(t) \propto \exp \left( Ht \right)$. However realistic
models of inflation usually deviate from this idealized situation,
since inflation must come to an end to allow the successes of the
standard Hot Big Bang to be reproduced after one second or so.

\subsection{Scalar fields and their potentials}

In order to obtain the required equation of state, a suitable material
must come to dominate the density of the Universe. Such a material is
a scalar field, which in particle physics is used to represent
spin-zero particles and which we represent by $\phi$ throughout. It
transforms as a scalar (that is, it is unchanged) under coordinate
transformations. In a homogeneous Universe, the scalar field is a
function of time alone.

At present no fundamental scalar field has been observed, but they
proliferate in modern particle physics theories. In particular,
supersymmetry associates a boson with every fermion (and vice versa),
giving a multitude of scalar fields in any supersymmetric theory
containing the Standard Model of particle physics.

The traditional starting point for particle physics models is the
action, which is an integral of the Lagrange density over space and
time and from which the equations of motion can be obtained. As an
intermediate step, one might write down the energy--momentum tensor,
which sits on the right-hand side of Einstein's equations. Rather than
begin there, I will take as my starting point expressions for the
effective energy density and pressure of a homogeneous scalar
field. These are obtained by comparison of the energy--momentum tensor
of the scalar field with that of a perfect fluid, and are
\begin{eqnarray}
\label{arl:effrho}
\rho_{\phi} & = & \frac{1}{2} \dot{\phi}^2 + V(\phi) \, \\
\label{arl:effp}
p_{\phi} & = & \frac{1}{2} \dot{\phi}^2 - V(\phi) \,.
\end{eqnarray}
One can think of the first term in each as a kinetic energy, and the
second as a potential energy. The potential energy $V(\phi)$ can be
thought of as a form of `configurational' or `binding' energy; it
measures how much internal energy is associated with a particular
field value (including the mass--energy of the particle number density
it represents). Normally, like all systems, scalar fields try to
minimize this energy; however, a crucial ingredient which allows
inflation is that scalar fields are not always very efficient at
reaching this minimum energy state.

Note in passing that a scalar field cannot in general be described by
an equation of state; there is no unique value of $p$ that can be
associated with a given $\rho$ as the energy density can be divided
between potential and kinetic energy in different ways. This is not
particularly significant for early Universe inflation, but will be
later when we discuss the present Universe.

\subsection{Models of inflation}

At present, understanding of fundamental physics is insufficient to
give clear guidance as to how to build inflationary models. The
present approach is therefore more phenomenological; we construct
models of inflation, develop their predictions, and ultimately compare
to observations in order to determine which properties are associated
with successful models. So far, this approach has narrowed the range
of possible models only modestly, and indeed theoretical ingenuity is
creating new models more rapidly than observational improvements are
ruling models out. Fortunately, it is projected that this state of
affairs will soon change; upcoming observations of microwave
anisotropies should have the capacity to exclude either all or nearly
all of existing inflationary models.

A model of inflation consists of some number of scalar fields, plus a
form for the potential of those fields.\footnote{It may also include
such complexities as deviations from general relativity (for example
as in scalar--tensor theories or the currently-popular braneworld
scenario), extra dimensional physics, etc, which I will not explore
here.} It may also require a specification of the means for ending
inflation. Ordinarily, the working assumption is that only a single
scalar field is dynamically important during inflation, possibly with
a second static field providing an additional contribution to the
energy density (see the later discussion of hybrid inflation). This
single-field paradigm is the simplest assumption that can be made, and
a useful initial goal is to investigate how well such a model can be
constrained by data, and indeed whether or not the entire class of
single-field models can be excluded. In the latter eventuality, the
question will arise as to whether a more complicated inflationary
model can keep the theory alive, or if one has to abandon the
inflationary model for structure formation altogether. I will say a
little on this later, but for the most part this article restricts its
attention to the single-field case.

Assuming that a single field $\phi$ gives the complete dynamics, the
model is given by a choice of $V(\phi)$. Until recently, this
potential would have been required to vanish at the minimum, in order
not to generate an unfeasibly large cosmological constant today (we
see from the effective density and pressure of a scalar field that a
non-vanishing potential at the minimum gives the equation of state $p
= - \rho$ mimicking a cosmological constant). In such a case,
inflation will necessarily terminate as the field approaches the
minimum.  However, this assumption need not be made if a second static
field supplies a contribution to the energy density, which may mean
that the global potential minimum occurs for a value of this field
differing from the one it has during inflation. It may therefore be
necessary to specify the value $\phi_{{\rm end}}$ at which inflation
terminates, as well as the potential.

\subsection{Equations of motion and solutions}

Let us assume there is only a single dynamical scalar field, and that
any extra contribution to the energy density from a second field is
included in its potential $V(\phi)$. The equations for an expanding
Universe containing a homogeneous scalar field are easily obtained by
substituting Eqs.~(\ref{arl:effrho}) and (\ref{arl:effp}) into the
Friedmann and fluid equations, giving
\begin{eqnarray}
\label{arl:scalfried}
H^2  =  \frac{8\pi}{3 \mpl^2} \left[ V(\phi) + \frac{1}{2} \dot{\phi}^2
	\right] \,, \\
\label{arl:scalwave}
\ddot{\phi} + 3 H \dot{\phi} = - V'(\phi) \,, \quad \quad
\end{eqnarray}
where prime indicates $d/d\phi$. Here I have ignored the curvature
term $k$, as it will quickly become negligible once inflation
starts. This is done for simplicity only; there is no obstacle to
including that term if one wished.

Since
\begin{equation}
\ddot{a} > 0 \Longleftrightarrow p < - \frac{\rho}{3}
       \Longleftrightarrow \dot{\phi}^2 < V(\phi)
\end{equation}
we will have inflation whenever the potential energy dominates. This
should be possible provided the potential is flat enough, as the
scalar field would then be expected to roll slowly. 

The standard strategy for solving these equations is the {\bf
slow-roll approximation} (SRA); this assumes that a term can be
neglected in each of the equations of motion to leave the simpler set
\begin{eqnarray}
H^2 & \; \simeq \; & \frac{8\pi}{3 \mpl^2} \, V \\
3 H \dot{\phi} & \; \simeq \; & -V'
\end{eqnarray}
If we define {\bf slow-roll parameters}~\cite{LL}
\begin{equation}
\label{arl:SR}
\epsilon(\phi) = \frac{\mpl^2}{16 \pi} \, \left( \frac{V'}{V}
	\right)^2 \quad ; \quad \eta(\phi) = \frac{\mpl^2}{8\pi}
	\, \frac{V''}{V} \,,
\end{equation}
where the first measures the slope of the potential and the second the
curvature, then necessary conditions for the slow-roll approximation
to hold are\footnote{Note that $\epsilon$ is positive by definition,
whilst $\eta$ can have either sign.}
\begin{equation}
\epsilon \ll 1 \quad ; \quad |\eta| \ll 1 \,.
\end{equation}
Unfortunately, although these are necessary conditions for the
slow-roll approximation to hold, they are not sufficient, since even
if the potential is very flat it may be that the scalar field has a
large velocity. A more elaborate version of the SRA exists, based on
the Hamilton--Jacobi formulation of inflation \cite{SB}, which is
sufficient as well as necessary \cite{LPB}.

Note also that the SRA reduces the order of the system of equations by
one, and so its general solution contains one less initial condition.
It works only because one can prove~\cite{SB,LPB} that the solution to
the full equations possesses an attractor property, eliminating the
dependence on the extra parameter.

\subsection{The relation between inflation and slow-roll}

\label{arl:infsr}

As it happens, the applicability of the slow-roll condition is closely
connected to the condition for inflation to take place, and in many
contexts the conditions can be regarded as equivalent. Let's quickly
see why.

The inflationary condition $\ddot{a} > 0$ is satisfied for a much
wider range of behaviours than just (quasi-)exponential expansion. A
classic example is power-law inflation $a \propto t^p$ for $p>1$,
which is an exact solution for an exponential potential
\begin{equation}
\label{exppot}
V(\phi) = V_0 \exp \left[ - \sqrt{\frac{16 \pi}{p}} \, \frac{\phi}{\mpl}
	\right] \,.
\end{equation}

We can rewrite the condition for inflation as
\begin{equation}
\frac{\ddot{a}}{a} = \dot{H} + H^2 > 0 \quad \Longleftrightarrow \quad
-\frac{\dot{H}}{H^2} < 1 \quad \stackrel{{\mbox{{\large
$\sim$}}}}{\Longleftrightarrow} \quad \frac{\mpl^2}{16\pi} \, \left(
\frac{V'}{V} \right)^2 < 1
\end{equation}
where the last manipulation uses the slow-roll approximation. The
final condition is just the slow-roll condition $\epsilon < 1$, and
hence
\[
\mbox{Slow-roll} \Longrightarrow \mbox{Inflation}
\]
Inflation will occur when the slow-roll conditions are satisfied
(subject to some caveats on whether the `attractor' behaviour has been
attained \cite{LPB}).

However, the converse is not strictly true, since we had to use the
SRA in the derivation. However, in practice
\begin{eqnarray}
\mbox{Inflation} & \quad \stackrel{{\mbox{{\large
	$\sim$}}}}{\Longrightarrow} \quad & \epsilon < 1 \nonumber \\
	\mbox{Prolonged inflation} & \quad \stackrel{{\mbox{{\large
	$\sim$}}}}{\Longrightarrow} \quad & \eta < 1 \nonumber
\end{eqnarray}
The last condition arises because unless the curvature of the
potential is small, the potential will not be flat for a wide enough
range of $\phi$.

\subsection{The amount of inflation}

The amount of inflation is normally specified by the logarithm of the
amount of expansion, {\em the number of e-foldings} $N$, given by
\begin{equation}
N \equiv \ln \frac{a(t_{{\rm end}})}{a(t_{{\rm initial}})}  = 
	\int_{t_{{\rm i}}}^{t_{{\rm e}}} \, H \, dt 
 \simeq  - \frac{8\pi}{\mpl^2} \int_{\phi_{{\rm i}}}^{\phi_{{\rm e}}}
 	\, \frac{V}{V'} \, d\phi \,, \label{efold}
\end{equation}
where the final step uses the SRA. Notice that the amount of inflation
between two scalar field values can be calculated without needing to
solve the equations of motion, and also that it is unchanged if one
multiplies $V(\phi)$ by a constant.

The minimum amount of inflation required to solve the flatness and
horizon problems is about 70 $e$-foldings, i.e.~an expansion by a
factor of about $10^{30}$. Although this looks large, inflation is
typically so rapid that most inflation models give much more.

\subsection{A worked example: polynomial chaotic inflation}

\label{arl:quadratic}

The simplest inflation model~\cite{Linbook} arises when one chooses a
polynomial potential, such as that for a massive but otherwise
non-interacting field, $V(\phi) = m^2 \phi^2/2$ where $m$ is the mass
of the scalar field. With this potential, the slow-roll equations are
\begin{equation}
3H\dot{\phi} + m^2 \phi = 0 \quad ; \quad H^2 = \frac{4\pi m^2
	\phi^2}{3\mpl^2} \,,
\end{equation}
and the slow-roll parameters are
\begin{equation}
\epsilon = \eta = \frac{\mpl^2}{4 \pi \phi^2} \,.
\end{equation}
So inflation can proceed provided $|\phi| > \mpl/\sqrt{4\pi}$, i.e.~as
long as we are not too close to the minimum.

The slow-roll equations are readily solved to give
\begin{eqnarray}
\phi(t) & = & \phi_{{\rm i}} - \frac{m \, \mpl}{\sqrt{12\pi}} \, t \,, \\
a(t) & = & a_{{\rm i}} \exp \left[ \sqrt{\frac{4\pi}{3}} \, \frac{m}{\mpl}
	\, \left( \phi_{{\rm i}} t - \frac{m \, \mpl}{\sqrt{48\pi}} t^2
	\right) \right] \,,
\end{eqnarray}
(where $\phi = \phi_{{\rm i}}$ and $a = a_{{\rm i}}$ at $t = 0$) and the 
total amount of inflation is
\begin{equation}
\label{arl:quadefold}
N_{{\rm tot}} = 2 \pi \, \frac{\phi_{{\rm i}}^2}{\mpl^2} - \frac{1}{2}
\,.
\end{equation}
This last equation can be obtained from the solution for $a$, but in
fact is more easily obtained directly by integrating
Eq.~(\ref{efold}), for which one needn't have bothered to solve the
equations of motion.

In the older inflationary literature, it was typically assumed that in
order for classical physics to be valid, we would require $V \ll
\mpl^4$ but no other restriction would be necessary. In particular,
considering scalar field theory in flat space-time, there is no
particular meaning to the actual value of the scalar field (which for
instance could be shifted by a constant). With those presumptions, we
see that while sufficient inflation requires $|\phi|> m_{{\rm Pl}}$,
one can readily get enough inflation provided $m$ is small enough, and
indeed we will see later that $m$ is in fact required to be small from
observational limits on the size of density perturbations produced,
allowing far more than the minimum amount of inflation required to
solve the various cosmological problems we originally set out to
solve.

\subsubsection{Multi-field theories}

While much of early investigation of inflation featured potentials
such as the massive field discussed above, such models are widely
regarded by inflation model builders as unsatisfactory. The reason is
that current thinking in particle physics is dominated by
supersymmetry, implying that in a cosmological context we should be
operating within the framework of supergravity. Once supergravity is
brought into play, the numerical value of the scalar field acquires a
well-defined meaning, and it is believed that its value must be less
than the (reduced) Planck mass $m_{{\rm Pl}}/\sqrt{8\pi}$ if the
potential is not to be vulnerable to large nonrenormalizable
corrections, which typically will destroy slow-roll and anyway will
render theoretical calculations unreliable. As we have just seen, one
cannot obtain sufficient inflation (or indeed any at all) with the
polynomial potential under this restriction, and this conclusion is
fairly generic for models where there is only a single scalar field.

\begin{figure}[tb]
\centering 
\leavevmode\epsfysize=6.5cm \epsfbox{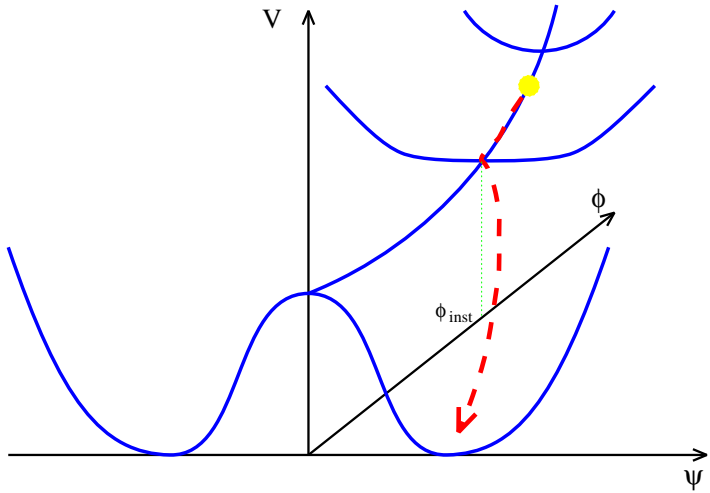}\\ 
\caption[hybridpot]{\label{arl:hybridpot} The potential for the hybrid
inflation model. The field rolls down the channel at $\psi = 0$ until
it reaches the critical $\phi$ value, then falls off the side to the
true minimum at $\phi = 0$ and $\psi = \pm M$.}
\end{figure}

An attractive way of circumventing this problem is the hybrid
inflation model \cite{hybrid}, where a second field provides an
additional energy density which dominates over that from the inflaton
itself. A typical potential takes the form
\begin{equation}
V(\phi,\psi) = \frac{\lambda}{4} \left( \psi - M \right)^4 + V(\phi) +
	\frac{1}{2}g^2 \psi^2 \phi^2 \,,
\end{equation}
where $g^2$ is the coupling constant governing the interaction between
the two fields. This is shown in Fig.~\ref{arl:hybridpot}. For large
inflaton values the coupling stabilizes $\psi$ at zero, where it
contributes a potential energy $\lambda M^4/4$ but otherwise does not
participate in the dynamics, so that the inflaton sees a potential
\begin{equation}
V_{{\rm eff}}(\phi) = V(\phi) + \frac{\lambda}{4} M^4 \,.
\end{equation}
The interesting case is where the constant term dominates, as it
provides extra friction to the $\phi$ equation of motion which makes
it roll much more slowly, enabling sufficient inflation without
violating the condition $\phi \ll m_{{\rm Pl}}/\sqrt{8\pi}$. Inflation
ends when the $\phi$ field drops below a critical value, destabilizing
the $\psi$ field and allowing the system to rapidly evolve into its
true minimum at $\phi = 0$, $\psi = \pm M$.

The original models \cite{hybrid} assumed that the inflaton potential
$V(\phi)$ was just that of a massive field, but unfortunately this
choice is vulnerable to large loop corrections which dominate over the
mass term. However many other possible models have been derived within
the hybrid framework; for an extensive discussion of this and other
model-building issues see the review of Lyth and Riotto \cite{LR}.

\subsection{Reheating after inflation}

During inflation, all matter except the scalar field (usually called
the inflaton) is redshifted to extremely low densities. {\bf
Reheating} is the process whereby the inflaton's energy density is
converted back into conventional matter after inflation, re-entering
the standard big bang theory. This may happen slowly due to decays of
individual inflaton particles, or may begin explosively due to a
coherent decay of the many-particle inflaton state. This latter
possibility is known as preheating \cite{KLS,preheat} and may convert
the bulk of the inflaton's energy density. Preheating will however
necessarily end before all the inflaton energy density is converted.

In the single field paradigm presently under discussion, it does not
really matter how reheating takes place, in the sense that the details
of the mechanism will not affect the predictions for large-scale
structure. I therefore won't discuss it further here, except to note
that the situation can be considerably more complicated in multi-field
theories. Then the details of reheating can be important for
determining the magnitude of density perturbations in the Universe,
and one also must consider possibilities such as that one or more of
the fields need not decay at all, but may instead survive up to the
present to act as dark matter or dark energy.

\section{Density Perturbations and Gravitational Waves}

By far the most important property of inflationary cosmology is that
it produces perturbations, in the form of both density perturbations
and gravitational waves. The density perturbations may be responsible
for the formation and clustering of galaxies, as well as creating
anisotropies in the microwave background radiation. The gravitational
waves do not affect the formation of galaxies, but may
contribute extra microwave anisotropies on the large angular scales
sampled by the COBE satellite \cite{COBE1,COBE4}. An alternative
terminology for the density perturbations is scalar perturbations and
for the gravitational waves is tensor perturbations, the terminology
referring to their transformation properties.

In this article I will focus on the nature of the predictions from the
inflationary cosmology rather than detailed comparison with
observations, which was the focus for a series of lectures at this
Summer School by Pedro Viana \cite{Pedro}.

\subsection{Production during inflation}

The ability of inflation to generate perturbations on large scales
comes from the unusual behaviour of the Hubble length during
inflation, namely that (by definition) the comoving Hubble length
decreases. When we talk about large-scale structure, we are primarily
interested in comoving scales, as to a first approximation everything
is dragged along with the expansion.  The qualitative behaviour of
irregularities is governed by their scale in comparison to the
characteristic scale of the Universe, the Hubble length.

In the big bang Universe the comoving Hubble length is always
increasing, and so all scales are initially much larger than it, and
hence unable to be affected by causal physics. Once they become
smaller than the Hubble length, they remain so for all time. In the
standard scenarios, COBE sees perturbations on large scales at a time
when they were much bigger than the Hubble length, and hence no
mechanism could have created them.

\begin{figure}[t!]
\centering 
\leavevmode\epsfysize=12cm \epsfbox{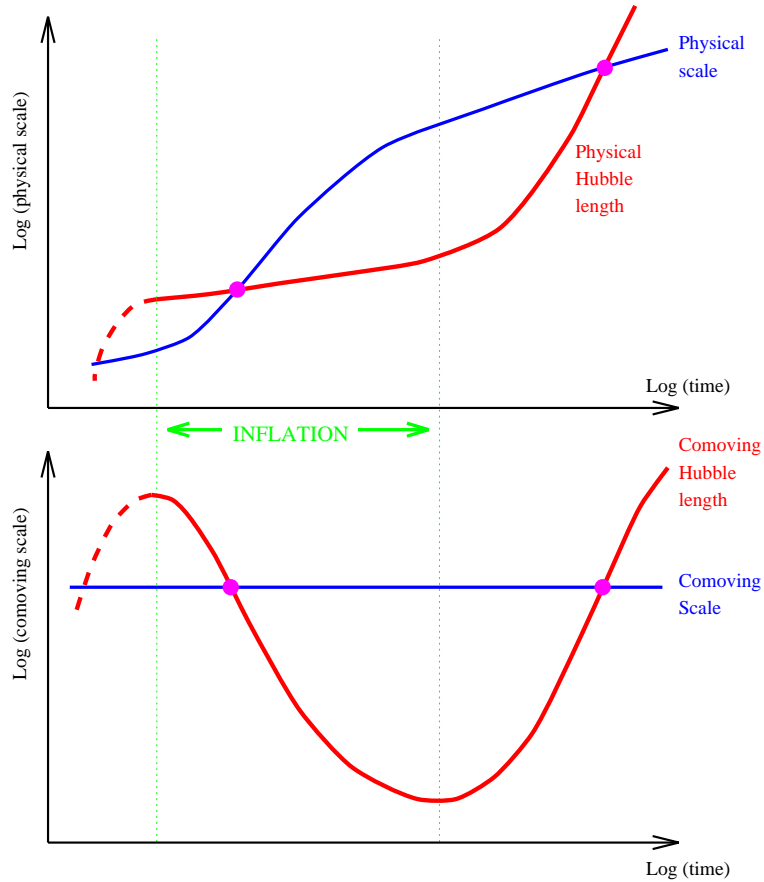}\\ 
\caption[scales]{\label{arl:scales} The behaviour of a given comoving
scale relative to the Hubble length, both during and after inflation,
shown using physical coordinates (upper panel) and comoving ones
(lower panel).}
\end{figure} 

Inflation reverses this behaviour, as seen in Figure~\ref{arl:scales}.
Now a given comoving scale has a more complicated history. Early on in
inflation, the scale could be well inside the Hubble length, and hence
causal physics can act, both to generate homogeneity to solve the
horizon problem and to superimpose small perturbations. Some time
before inflation ends, the scale crosses outside the Hubble radius
(indicated by a circle in the lower panel of Figure~\ref{arl:scales})
and causal physics becomes ineffective. Any perturbations generated
become imprinted, or, in the usual terminology, `frozen in'. Long
after inflation is over, the scales cross inside the Hubble radius
again.  Perturbations are created on a very wide range of scales, but
the most readily observed ones range from about the size of the
present Hubble radius (i.e.~the size of the presently observable
Universe) down to a few orders of magnitude less. On the scale of
Figure~\ref{arl:scales}, all interesting comoving scales lie extremely
close together, and cross the Hubble radius during inflation very
close together.

It's all very well to realize that the dynamics of inflation permits
perturbations to be generated without violating causality, but we need
a specific mechanism. That mechanism is quantum
fluctuations. Inflation is trying as hard as it can to make the
Universe perfectly homogeneous, but it cannot defeat the Uncertainty
Principle which ensures that there are always some irregularities left
over. Through this limitation, it is possible for inflation to
adequately solve the homogeneity problem and in addition leave enough
irregularities behind to attempt to explain why the present Universe
is not completely homogeneous.

The size of the irregularities depends on the energy scale at which
inflation takes place. It is outside the scope of these lectures to
describe in detail how this calculation is performed; the interested
reader can find the full details in Ref.~\cite{LLbook}. I
will just quote the results, which we can go on to apply.

The formulae for the amplitude of density perturbations, which I'll
call $\delh(k)$, and the gravitational waves, $A_{{\rm G}}(k)$,
are~\footnote{The precise normalization of the spectra is arbitrary,
as are the number of powers of $k$ included. I've made my favourite
choice here (following Refs.~\cite{LLbook,LLKCBA}), but whatever
convention is used the normalization factor will disappear in any
physical answer. For reference, the usual power spectrum $P(k)$ is
proportional to $k \delh^2(k)$.}
\begin{eqnarray}
\label{arl:delh}
\delh(k) & = & \left. \sqrt{\frac{512 \pi}{75}} \, \frac{V^{3/2}}{\mpl^3
	|V'|} \right|_{k = aH} \,, \\
\label{arl:AG}
A_{{\rm G}}(k) & = & \left. \sqrt{\frac{32}{75}} \, \frac{V^{1/2}}{\mpl^2}
	\right|_{k = aH} \,.
\end{eqnarray}
Here $k$ is the comoving wavenumber; the perturbations are normally
analyzed via a Fourier expansion into comoving modes. The power
spectra $\delh(k)$ and $A_{{\rm G}}(k)$ measure the typical size of
perturbations on a scale $k$.  The right-hand sides of the above
equations are to be evaluated at the time when $k = aH$ during
inflation, which for a given $k$ corresponds to some particular value
of $\phi$. We see that the amplitude of perturbations depends on the
properties of the inflaton potential at the time the scale crossed the
Hubble radius during inflation. The relevant number of $e$-foldings
from the end of inflation is given by~\cite{LLrep}
\begin{equation}
N \simeq 62 - \ln \frac{k}{a_0 H_0} + \mbox{numerical correction} \,,
\end{equation}
where `numerical correction' is a typically smallish (order a few)
number which depends on the energy scale of inflation, the duration of
reheating and so on. Normally it is a perfectly fine approximation to
say that the scales of interest to us crossed outside the Hubble
radius 60 $e$-foldings before the end of inflation. Then the
$e$-foldings formula
\begin{equation}
\label{arl:efold2}
N \simeq - \frac{8\pi}{\mpl^2} \int_{\phi}^{\phi_{{\rm end}}} \, 
	\frac{V}{V'} \, d\phi \,,
\end{equation}
tells us the value of $\phi$ to be substituted into
Eqs.~(\ref{arl:delh}) and (\ref{arl:AG}).

\subsection{A worked example}

The easiest way to see what is going on is to work through a specific
example, the $m^2 \phi^2/2$ potential which we already saw in
Section~\ref{arl:quadratic}. We'll see that we don't even have to
solve the evolution equations to get our predictions.
\begin{enumerate}
\item Inflation ends when $\epsilon = 1$, so $\phi_{{\rm end}} \simeq
\mpl/\sqrt{4\pi}$.
\item We're interested in 60 $e$-foldings before this, which from
Eq.~(\ref{arl:quadefold}) gives $\phi_{60} \simeq 3 \mpl$.
\item Substitute this in:
\[
\delh \simeq 12 \, \frac{m}{\mpl} \quad ; \quad A_{{\rm G}} \simeq 1.4
	\, \frac{m}{\mpl} 
\]
\item Reproducing the COBE result requires~\cite{BLW} 
$\delh \simeq 2 \times 10^{-5}$ 
(provided $A_{{\rm G}} \ll \delh$), so we need $m \simeq 10^{-6} 
\mpl$.
\end{enumerate}

Because the required value of $m$ is so small, that means it is easy
to get sufficient inflation to solve the cosmological problems if one
only requires the classicality condition $V< \mpl^4$. Since that
condition implies only that $\phi < \mpl^2/m \simeq 10^6 \mpl$, and as
$N_{{\rm tot}} \simeq 2\pi \phi^2/\mpl^2$, we can get up to about
$10^{13}$ $e$-foldings in principle. This compares extremely
favourably with the 70 or so actually required.

\subsection{Observational consequences}

Observations have moved on beyond us wanting to know the overall
normalization of the potential. The interesting things are
\begin{enumerate}
\item The scale-dependence of the spectra.
\item The relative influence of the two spectra.
\end{enumerate}
These can be neatly summarized using the slow-roll parameters $\epsilon$
and $\eta$ we defined earlier \cite{LL}.

The standard approximation used to describe the spectra is the {\bf
power-law approximation}, where we take
\begin{equation}
\delh^2(k) \propto k^{n-1} \quad ; \quad A_{{\rm G}}^2(k) 
	\propto k^{n_{{\rm G}}} \,,
\end{equation}
where the spectral indices $n$ and $n_{{\rm G}}$ are given by
\begin{equation}
n-1 = \frac{d \ln \delh^2}{d \ln k} \quad ; \quad n_{{\rm G}} =
	\frac{d \ln A_{{\rm G}}^2}{d \ln k} \,.
\end{equation}
The power-law approximation is usually valid because only a limited
range of scales are observable, with the range $1$ Mpc to $10^4$ Mpc
corresponding to $\Delta \ln k \simeq 9$.

The crucial equation we need is that relating $\phi$ values to when a scale 
$k$ crosses the Hubble radius, which from Eq.~(\ref{arl:efold2}) is
\begin{equation}
\frac{d \ln k}{d \phi} = \frac{8\pi}{\mpl^2} \, \frac{V}{V'} \,.
\end{equation}
(since within the slow-roll approximation $k \simeq \exp N$). Direct
differentiation then yields~\cite{LL}
\begin{equation}
n = 1 - 6\epsilon + 2 \eta \quad ; \quad n_{{\rm G}} = -2 \epsilon \,,
\end{equation}
where now $\epsilon$ and $\eta$ are to be evaluated on the appropriate
part of the potential.

Finally, we need a measure of the relevant importance of density
perturbations and gravitational waves. The natural place to look is
the microwave background; a detailed calculation which I cannot
reproduce here (see e.g.~Ref.~\cite{LLrep}) gives
\begin{equation}
\label{arl:relgrav}
R \equiv \frac{C_{\ell}^{{\rm GW}}}{C_{\ell}^{{\rm DP}}} \simeq \,
	4 \pi \epsilon \,.
\end{equation}
Here the $C_{\ell}$ are the contributions to the microwave multipoles,
in the usual notation.\footnote{Namely, $\Delta T/T = \sum a_{\ell m}
Y^{\ell}_{m}(\theta,\phi)$, $C_{\ell} = \langle | a_{\ell m}|^2
\rangle$.}

From these expressions we immediately see
\begin{itemize}
\item If and only if $\epsilon \ll 1$ and $|\eta| \ll 1$ do we get $n
\simeq 1$ and $R \simeq 0$.
\item Because the coefficient in Eq.~(\ref{arl:relgrav}) is so large,
gravitational waves can have a significant effect even if $\epsilon$
is quite a bit smaller than one.
\end{itemize}

At present, a large number of inflationary models exist covering a
large part of the $n$--$r$ parameter space. Observations are just
beginning to narrow down the allowed region, and in the future
satellite microwave anisotropy experiments such as MAP and Planck
\cite{sats} should determine $n$ sufficiently accurately to exclude
almost all models of inflation on that basis, and may be able to
measure $r$ as well.

The principal observational challenge is to untangle the effects of
the inflationary parameters ($\delh(k_0)$, $n$ and $r$) from all the
other parameters required to specify a complete cosmological model,
such as the Hubble constant, the density of each component of matter,
and so on. The two sets of parameters cannot be studied separately; an
attempt to match the observations must fit for both simultaneously. A
typical set of parameters likely to be important in determining
predictions for observations such as microwave anisotropies contains
about ten different parameters, with some authors suggesting this list
extends up to fifteen or more. It is a testament to the predicted
accuracy of upcoming observations that considerable progress is
expected in this direction over the next decade.

\subsection{Testing the idea of inflation}

An extremely useful sentence to bear in mind in considering how to
test the inflationary paradigm is the following:
\begin{quote}
The {\em simplest} models of inflation predict a
\underline{spatially-flat Universe} containing \underline{gaussian
distributed} \underline{tensor} and \underline{adiabatic scalar}
perturbations, which are in their \underline{growing mode} with
\underline{almost power-law spectra}.
\end{quote}
The underlined phrases indicate the characteristic predictions, while
the emphasised word `simplest' stresses that these predictions are not
all generic. With the possible exception of the power-law form, the
class of simplest models encompasses all the single-field models, plus
many other models which prove to be dynamically equivalent to them

So far, all these characteristic predictions has been borne out,
though the strength of the tests differs \cite{Pedro}. Only spatial
flatness has really been confirmed at a convincing level, and even
that only very recently \cite{Boom,Max}. The others are promising
hypotheses which are part of the currently-favoured view as to how
structure arose.

The immediate goal is to test these hypotheses, and if they remain
valid to use measured quantities such as $n$ and $r$ to establish
which subset of the simplest models best fits the data. Our current
aim, therefore, is to test the simplest models of inflation. If they
are found wanting, then the way in which they fail will be indicative
of whether it is worthwhile to study the wider range of inflation
models, or if attention is best refocussed elsewhere. One should also
stress that the aim is to test inflation as the {\em sole} origin of
structure; one can consider admixtures of the inflationary
perturbations with those of another source; this is fine if positive
evidence is forthcoming but it will of course be impossible ever to
exclude the possibility of an admixture at some level.

\subsubsection{Growing mode perturbations}

In general perturbations evolving on scales larger than the horizon
have both growing and decaying modes, which on horizon entry become
the baryonic oscillations. However because inflationary perturbations
were laid down in the distant past, they evolve to become completely
dominated by the growing mode by the time they enter the horizon. This
leads to a fixed phase of oscillations for the baryons on a given
scale; for example, at any given time (such as decoupling) there are
particular scales on which the oscillations are at zero amplitude. On
those scales the corresponding microwave anisotropies will be at their
smallest.  This contrasts with topological defect scenarios, where the
modes are sourced after entering the horizon and in general have a
mixture of the two perturbation modes, meaning that the phase of
oscillation is not determined.

The growing mode prediction is perhaps the most important generic
prediction of inflation, and one that cannot be avoided. It leads to
characteristic predictions; the phase coherence of the oscillations
that results is what leads to the oscillatory structure in the
microwave anisotropy spectrum \cite{coherence}. Such oscillations will
arise whether the inflationary perturbations are gaussian or
non-gaussian, and whether they are adiabatic or non-adiabatic. The
prediction of an oscillatory structure in the microwave anisotropies
is therefore a key testable prediction of inflation. That is not to
say that the discovery of such would prove inflation, as other
mechanisms may prove capable of also creating oscillations. But if the
oscillations are not seen, the paradigm of inflation as the sole
origin of structure will be excluded. The combined BOOMERanG and
Maxima data are not quite sufficient for a definitive verdict, but
tests of this prediction are imminent.

\subsubsection{Gaussianity and adiabaticity}

While gaussianity and adiabaticity are predictions of the simplest
models, it is well advertised that there exist inflationary models
which violate these hypotheses. Whether observations violating
gaussianity or adiabaticity would exclude inflation depends on the
type observed. For example, chi-squared distributed perturbations are
relatively easy to generate from inflation, by ensuring that the
leading contribution to perturbations is the square of a gaussian
field. On the other hand, it would be pointless to try and find an
inflation model if line discontinuities in the microwave sky were
discovered.

\subsubsection{Tensor and vector modes}

Large-scale tensor modes are a generic prediction of inflation, but
unfortunately the amplitude depends on the inflation model and there
is no automatic implication that the tensor modes will have sufficient
amplitude to be detectable by forseeable technology (indeed if
anything theoretical prejudice suggests the opposite
\cite{lythgw}). If tensor modes are discovered, for example by the
Planck satellite, that will be exceedingly strong support for
inflation. However tensor modes do not formally constitute a test of
inflation, since their absence does not exclude the paradigm.

By contrast, the simplest inflationary models do not produce vector
perturbations. If those are seen, then at the very least it will
represent a considerable challenge to model-builders to find a way of
accounting for them from within the inflationary paradigm.

\subsubsection{Power-law spectra}

It is currently fashionable to assume that the perturbation spectra,
especially the scalars, take on a power-law form, just as it was
fashionable pre-COBE to assume the Harrison--Zel'dovich form. Both
these cases are approximations to the true situation, and as
observations improve the power-law approximation may be found wanting
just as the Harrison--Zel'dovich one was post-COBE.  Staying within
the single-field paradigm, unless prominent features are dialed into
the inflationary potential, deviations from power-law behaviour are
however predicted to be small, perhaps even by the standards to be
imposed by the Planck satellite. Nevertheless, one should test for
deviations when improved data come in \cite{recon,KosT,CGL}.

\section{Acceleration in the present Universe}

\subsection{Cosmological constant: exact or effective?}

The discovery of strong evidence that the present Universe is
accelerating is one of the most striking discoveries of recent
cosmology. The results from Type Ia supernovae have been confirmed by
two separate teams \cite{Sn}, and have subsequently stood up to
detailed examination of the underlying mechanisms. The general
conclusion has been further bolstered by support \cite{Boom,Max} from
the microwave background for the expectation that the Universe is
spatially flat, which coupled with the observed low matter density
gives independent demonstration of the need for a cosmological
constant or similar.

These developments have confirmed the $\Lambda$CDM model as the
standard cosmological model. It has a matter density around one third
of the critical density, with the cosmological constant making up the
remainder. It explains a range of phenomena concerning the observed
content and dynamics of the Universe (including the age and the
observed acceleration), and provides a model of structure formation
which is not in serious disagreement with any existing
observations. It is accepted almost universally amongst cosmologists
that this model is currently the most observationally viable.

Despite that, the fraction of cosmologists willing to accept the
$\Lambda$CDM model is significantly less, because of serious
philosophical objections to the idea of a cosmological constant. First
of all, there is absolutely no fundamental understanding of the
cosmological constant. The standard interpretation is that it
corresponds to the energy density of the vacuum state, but attempts to
compute it typically result in answers so large as to be immediately
excluded, leading to the suspicion that some as-yet-unknown physical
mechanism sets it precisely to zero. This is known as the cosmological
constant problem. Secondly, that the cosmological constant has just
come to dominate at the present epoch marks out this time as a special
point in the Universe's history; at a redshift of a few it was
completely negligible, and within a Hubble time it will be completely
dominant. As the historical theme of cosmology has been to avoid
placing ourselves at preferred locations both spatially and
temporally, this situation causes great unease and has become known as
the coincidence problem.

The observed value of the cosmological constant is that it is close to
the critical density, which in particle physics units gives
\begin{equation}
\Lambda \simeq (0.003 \, {\rm eV})^4 \simeq 10^{-120} \,  m_{{\rm
Pl}}^4 \,.
\end{equation}
The na\"{\i}vest fundamental physics estimates of the vacuum energy
assume no suppression of contributions to the vacuum energy all the
way to the Planck scale, and thus yield $\Lambda \simeq m_{{\rm
Pl}}^4$. Reasonable arguments can bring this down to the scale of
supersymmetry, $\Lambda = m_{{\rm susy}}^4 \simeq 10^{-64} \, m_{{\rm
Pl}}^4$, by presuming that at energies where supersymmetry is restored
the bosonic and fermionic contributions to the cosmological constant
cancel exactly. The vast discrepancy between expectation and reality
puts us in an awkward situation; either we have to find an argument
for a dimensionless prefactor to that estimate making up the remaining
factor of $10^{-56}$, or we might conclude that there must be some
fundamental principle setting it to zero leaving us to look elsewhere.

For an early Universe cosmologist, the temptation is irresistible to
try and employ the same physical mechanisms in the present Universe as
we did for inflation in the early Universe. Both correspond to an
acceleration of Universe, and both could be the consequence of
domination of the Universe's energy density by the potential energy of
a scalar field. Because we know, for example from nucleosynthesis,
that there must have been a longer deceleration phase in between these
two epochs, with some notable exceptions \cite{quintinf} the standard
assumption is that different scalar fields are responsible for the two
epochs, but that the physical mechanism is analogous. Describing the
cosmological constant as an effective one via a scalar field has
become known as {\bf quintessence}, though that is a recent term to
describe an idea with quite a long history \cite{RP,Wett}.

As far as modelling is concerned, there are some differences between
the early Universe and today. In the early Universe, we know that
inflation has to come to an end, and arranging that this happens
satisfactorily is a significant constraint on model building. By
contrast, we have no idea whether the present acceleration will end in
the future. Secondly, in the present Universe we are interested in the
earliest stages of the acceleration, and indeed the way in which the
Universe entered the accelerating phase, and so we cannot neglect the
effect of the non-scalar field matter as is common in early Universe
inflation studies.

Broadly speaking, one can recognize three different possibilities. One
is that in our Universe the scalar field is in the true minimum of its
potential, whose value happens to be non-zero. Phenomenologically this
is no different from a true cosmological constant. Secondly, we might
live in a metastable false vacuum state, destined at some future epoch
to tunnel into the true vacuum. Such tunnelling would almost certainly
have drastic consequences for the material world, but we can be
somewhat reassured by the fact that the decay time is at least a
Hubble time. The false vacuum in this case also mimics a cosmological
constant. Neither of these scenarios give us the possibility of
addressing the coincidence problem, so mostly attention has been
focussed on the third possibility, that the scalar field is slowly
rolling in its potential akin to the chaotic inflation models. In this
situation, the effective cosmological constant is in fact not constant
at all but rather is slowly varying, and as such observations can seek
to distinguish it from a pure cosmological constant.

\subsection{Scaling solutions and trackers}

One reason for having optimism that quintessence can at least address
the coincidence problem comes from an interesting class of solutions
known as scaling solutions or trackers. These arise because the scalar
field does not have a unique equation of state $p \equiv
p(\rho)$. Although we can usefully define an effective equation of
state
\begin{equation}
w_\phi \equiv \frac{p_\phi}{\rho_\phi} = \frac{\dot{\phi}/2 -
V(\phi)}{\dot{\phi}/2 + V(\phi)} \,,
\end{equation}
in which terminology cosmological constant behaviour corresponds to $w
= -1$, the scalar field velocity depends on the Hubble expansion,
which in turn depends not only on the scalar field itself but on the
properties of any other matter that happens to be around. That is to
say, the scalar field responds to the presence of other matter.

A particularly interesting case is the exponential potential
\begin{equation}
V(\phi) = V_0 \exp \left[ - \sqrt{\frac{8 \pi}{m_{{\rm Pl}}^2}} \,
\lambda \phi \right] \,,
\end{equation}
which we already saw in the early Universe context as
Eq.~(\ref{exppot}).  If there is only a scalar field present, this
model has inflationary solutions $a \propto t^{2/\lambda^2}$ for
$\lambda^2 < 2$, and non-inflationary power-law solutions
otherwise. However, if we add conventional matter with equation of
state $p = w \rho$, a new class of solutions can arise, which turn out
to be attractors whenever they exist \cite{RP,Wett,scaling}. These
solutions take the form of scaling solutions, where the scalar field
energy density (indeed both its potential and kinetic energy density
separately) exhibit the same scaling with redshift as the conventional
matter. That is to say, the scalar field mimics whatever happens to be
the dominant matter in the Universe. So, for example, in a
matter-dominated Universe, we would find $\rho_\phi \propto 1/a^3$. If
the matter era were preceded by a radiation era, at that time the
scalar field would redshift as $1/a^4$, and it would make a smooth
transition between these behaviours at equality. The ratio of
densities is decided only by the fundamental parameters $\lambda$ and
$w$. So, at any epoch one expects the scalar field energy density to
be comparable to the conventional matter.

Unfortunately this is not good enough. We don't want the scalar field
to be behaving like matter at the present, since it is supposed to be
driving an acceleration, and we need it to be negligible in the fairly
recent past. This requires us to consider alternatives to the
exponential potential, a common example being the inverse power-law
potential \cite{RP}
\begin{equation}
V(\phi) = \lambda m_{{\rm Pl}}^4 \left( \frac{\phi}{m_{{\rm Pl}}}
\right)^{\alpha} \,,
\end{equation}
where $\alpha < 0$. In fact, Liddle and Scherrer \cite{LS} gave a
complete classification of Einstein gravity models with scaling
solutions, defined as models where the scalar field potential and
kinetic energies stay in fixed proportion. The exponential potential
is a particular case of that, but in general the scaling of the
components of the scalar field energy density need not be the same as
the scaling of the conventional matter, and indeed the inverse
power-law potential is an example of that; if the conventional matter
is scaling as $1/a^m$ where $m=3(1+w)$, there is an attractor solution
in which the scalar field densities will scale as
\begin{equation}
\rho_{{\rm kinetic}} \propto \rho_{{\rm potential}} \propto
\frac{1}{a^n} \quad ; \quad n = \frac{\alpha}{\alpha - 2} \, m \,.
\end{equation}
With negative $\alpha$, the scalar field energy density is redshifting
more slowly and eventually overcomes the conventional matter, at which
point the Universe starts to accelerate.

This type of scenario can give a model capable of explaining the
observational data, though it turns out that quite a shallow power-law
is required in order to get the field to be behaving sufficiently like
a cosmological constant (current limits require $w_\phi < -0.6$ at the
present epoch, where $\Omega_\phi \simeq 0.7$ \cite{wobs}). Also, the
epoch at which the field takes over and drives an acceleration is
still more or less being put in by hand; it turns out that the
acceleration takes over when $\phi \simeq m_{{\rm Pl}}$, and so
$\lambda \simeq 10^{-120}$ is required to ensure this epoch is delayed
until the present.

Various other forms of the potential have been experimented with, and
many possibilities are known to give viable evolution histories
\cite{models}. While such models do give a framework for interpretting
the type Ia supernova results, in many cases with the possibility
ultimately of being distinguished from a pure cosmological constant, I
believe it is fair to say that so far no very convincing resolution of
either the cosmological constant problem or the coincidence problem
has yet appeared. However, quintessence is currently the only approach
which has any prospect of addressing these issues.

\subsection{Challenges for quintessence}

At the moment quintessence is viable but lacks clear motivation, and
greater observational input is needed before it will be possible to
decide if it is really along the right track. The supernova
measurements so far are not capable of giving very serious
constraints, but this will hopefully change in the near future, both
by observations extending to higher redshift and by improved
measurements and statistics both at redshifts approaching one, where
most of the current data is, and in the local calibrating sample. One
exciting prospect is a dedicated satellite project which is currently
under consideration, which may allow current systematic errors to be
addressed as well as delivering a vastly larger data set. Concerning
non-supernova observations, improved constraints on parameters such as
$\Omega_0$, for instance from structure formation, would be welcome
and will inform studies of quintessence. It may even eventually prove
possible to spot the time evolution of quintessence by its effect on
the microwave background through the integrated Sachs--Wolfe effect.

The two outstanding issues are the magnitude of the cosmological
constant and why it should come to dominate at the present epoch. For
a pure cosmological constant these are one and the same, and at
present no lines of attack offer themselves. For quintessence, one
question is whether there is an energy scale in fundamental physics
comparable to the cosmological constant energy density (numerologists
might consider for instance the similarity between the cosmological
constant energy density and the measured neutrino mass-squared
difference). 

As to the coincidence problem, a very attractive, though as yet
unrealized, notion is triggered quintessence, where the scalar field
comes to dominate because a change in its behaviour is triggered
by some other event. The obvious such recent event is
matter--radiation equality, which occurred at a redshift of a few
thousand. The ideal scenario would have perfect tracking during
radiation domination, but a different behaviour during the matter era
leading to domination after a suitable delay. At first sight, it seems
that it might be possible to arrange this by coupling the quintessence
field to the Ricci scalar, which vanishes in a perfect radiation
dominated universe but is non-zero during matter
domination.\footnote{Actually typically $R$ is not zero during the
radiation era, because it still receives a contribution from the
subdominant matter which is actually larger than in the subsequent
matter-dominated era. But $R \ll H^2$ is a more significant criterion
which is true during the radiation era.} Several such models have been
studied \cite{exquint}, but unfortunately triggered quintessence has
yet to be realised in a natural way and it is looking increasingly
doubtful that this line of attack will prove fruitful.

\section{Summary}

The aim of this article has been to recount how we go about describing
two epochs of the Universe's evolution during which it may have
experienced accelerated expansion, and to highlight possible
similarities in those descriptions. It is an appealing possibility
that the same underlying mechanism may be responsible in each case.

Concerning the early Universe, I have introduced some of the facets of
inflation in a fairly simple manner. If you are interested in going
beyond this, a detailed account of all aspects of the inflationary
cosmology can be found in Ref.~\cite{LLbook}. Additional material on
particle physics and model building aspects of inflation can be found
in Ref.~\cite{LR}.

At present, inflation is the most promising candidate theory for the
origin of perturbations in the Universe. Different inflation models
lead to discernibly different predictions for these perturbations, and
hence high-accuracy measurements are able to distinguish between
models, excluding either all or the vast majority of them.  Since its
inception, the inflationary cosmology has been a gallery of different
models, and the gallery has continually needed extension after
extension to house new acquisitions. In all the time up to the
present, very few models have been discarded. However, the near future
holds great promise to finally begin to throw out inferior models,
and, if the inflationary cosmology survives as our model for the
origin of structure, we can hope to be left with only a narrow range
of models to choose between.

Concerning the present Universe, quintessence is an interesting idea
which is still in its early days as far as observations are
concerned. It provides a framework in which to study the possibility
of an accelerating Universe at present, but so far it has not
completely lived up to its initial promise as a way of challenging
problems such as the coincidence problem. There appears to be
plentiful scope for further investigations of possible quintessence
scenarios to try and remedy this shortcoming, especially in
anticipation of improved observations in years to come. At present,
there is little in the way of rivals to quintessence in terms of
allowing a quantitative description of the recent evolution of the
Universe. 

%%%%%%%%%%%%%%%%%%%%%%%%%%%%%%%%%%%%%%%%%%%%%%%%%%%%%%%%%%%%%%%%%%%%%%

%%%%%%%%%%%%%%%%%%%%%%%%%%%%%%%%%%%%%%%%%%%%%%%%%%%%%%%%%%%%%%%%%%%%%%

\end{document}